# FULLY INTEGRATED AUTOMATIC REUSABLE MICROFLUIDIC SETUP FOR IMMOBILIZATION, ANALYSIS AND NON-SELECTIVE RELEASE OF PARTICLES

Alexandre CHARGUERAUD[1, 2], Lars KOOL[3] and Jacques FATTACCIOLI[1, 2]

**AFFILIATIONS**

[1] Laboratoire CPCV, Département de Chimie, École Normale Supérieure, PSL Research University, Sorbonne Université, CNRS, 75005 Paris, France

[2] Institut Pierre-Gilles de Gennes pour la Microfluidique, 75005 Paris, France

[3] Plateforme Technologique, Institut Pierre-Gilles de Gennes pour la Microfluidique, CNRS UAR3750, ESPCI-PSL, 75005 Paris, France

Corresponding authors: jacques.fattaccioli@ens.psl.eu

# ABSTRACT

We present a microfluidic device that enables trapping, analysis and on-demand release of individual microparticles using membrane deformation induced by pneumatic actuation. Inspired by the *Pachinko*-like architecture, the system integrates an array of traps supported by a deformable membrane. The latter is actuated by pressure control in a pneumatic chamber bonded above it. Unlike conventional designs with rigid guiding structures, our modified configuration maximizes trapping membrane mobility by introducing a wide, elliptical cavity and suspended rectilinear flow guides.

To characterize the device, we performed deformation measurements under controlled negative pressure using confocal microscopy. These measurements revealed pressure-dependent displacement profiles, which were compared to several theoretical models. A strong agreement was observed with the thick-membrane approximation, confirming its suitability for future scaling. Deformation in the z-direction reached over 120 µm under –100 mbar in the pneumatic chamber, enabling the release of 80-µm-diameter particles.

Trapping and release sequences were demonstrated under a microscope by applying controlled fluidic and pneumatic pressures. The device reproducibly captured and released microparticles within seconds. To highlight the automation potential, we integrated a custom Python-based interface using µManager libraries to control imaging, motion, and fluidic systems. The platform autonomously performed full cycles of particle trapping, image acquisition, and release, including priming steps to prevent air bubble formation in the PDMS microchannels.

This work demonstrates a robust and fully automated microfluidic platform capable of reversible particle immobilization. The system is suitable for applications requiring high-throughput single-object analysis, such as screening assays, mechanobiology, or dynamic imaging workflows.

1. **Introduction**

Microfluidic trapping is a cornerstone of single-cell analysis, offering high-resolution control over individual biological particles. While numerous trapping strategies have been developed, efficient and controllable release mechanisms are less advanced. This limitation prevents the reuse of microfluidic chips across successive experiments, reducing their applicability in long-term, time-resolved studies.

Here, we present an automated single-cell analysis platform that enables successive trapping and release of particles within a reusable microfluidic chip. This approach supports high-throughput, repeatable analysis of large particle populations, expanding the capabilities of microfluidic systems for applications such as monitoring cellular responses over time.

Several strategies exist for particle trapping in microfluidics [1–4], including hydrodynamic confinement [5], dielectrophoresis [6], and structured trap arrays [7]. Among these, trap arrays are particularly effective for single-cell analysis, allowing passive capture of numerous particles without requiring external forces. However, most current designs lack integrated release functionality. While trap-and-release systems have been reported [8], they often suffer from low throughput and limited scalability, relying on sequential operation (e.g., first-in–last-out [9]) or one control valve per trap [10], making them unsuitable for high-content screening. In contrast, array-based systems can trap hundreds of particles simultaneously but rarely offer simultaneous and controlled release.

Material choice also plays a crucial role in microfluidic design [11]. Polydimethylsiloxane (PDMS) is widely used due to its favorable optical and mechanical properties [12]. Notably, its elasticity enables pneumatic deformation, which has been widely used in the development of Quake-style valves [13]. Prior studies have leveraged this property to modulate fluid flow [14] and, more recently, to explore its potential for coupling deformation with single-cell trapping [15,16]. Despite promising demonstrations, efficient and scalable implementations remain rare.

Enabling the release of trapped particles would allow repeated use of a single chip, particularly beneficial for monitoring reactions with slow kinetics. Despite the widespread use of Pachinko-like trap arrays, to our knowledge, no simple and effective method has yet been established to render them reusable.

In this work, we introduce a microfluidic platform that combines a deformable PDMS chip mounted on an interface for automated control and imaging. This system enables efficient trapping, full particle release, and reuse of the same microfluidic chip for long-term single-cell analysis in dynamically evolving environments (Fig. 1).

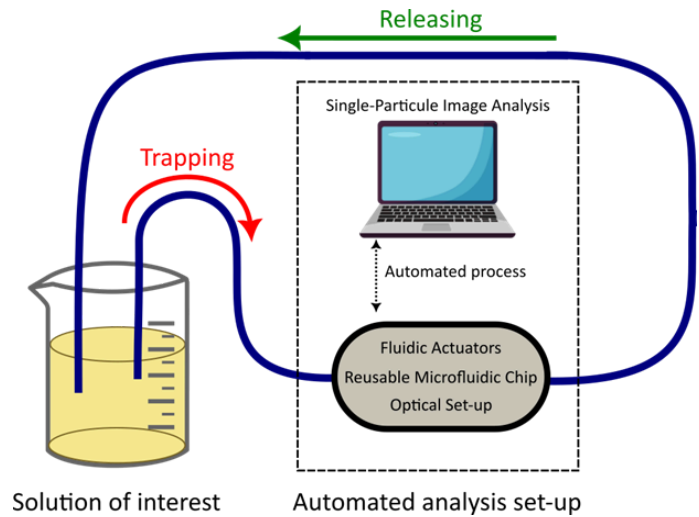

*Figure 1 – Schematic of the Automated Solution Analysis Device. This figure illustrates a use case involving a reusable trap array. A solution containing microparticles is analyzed by the device, which extracts and characterizes a significant number of particles before reinjecting them into the solution. This process can be repeated as many times as needed, depending on the experiment chosen by the user.*

## 2. Materials And Methods

### 2.1. Layout design and image analysis

Mask layouts were designed with AutoCAD software. Image processing and analysis were done with Fiji/ImageJ [17] and Python software. Data processing and analysis were performed with Python software.

### 2.2. Chip Fabrication

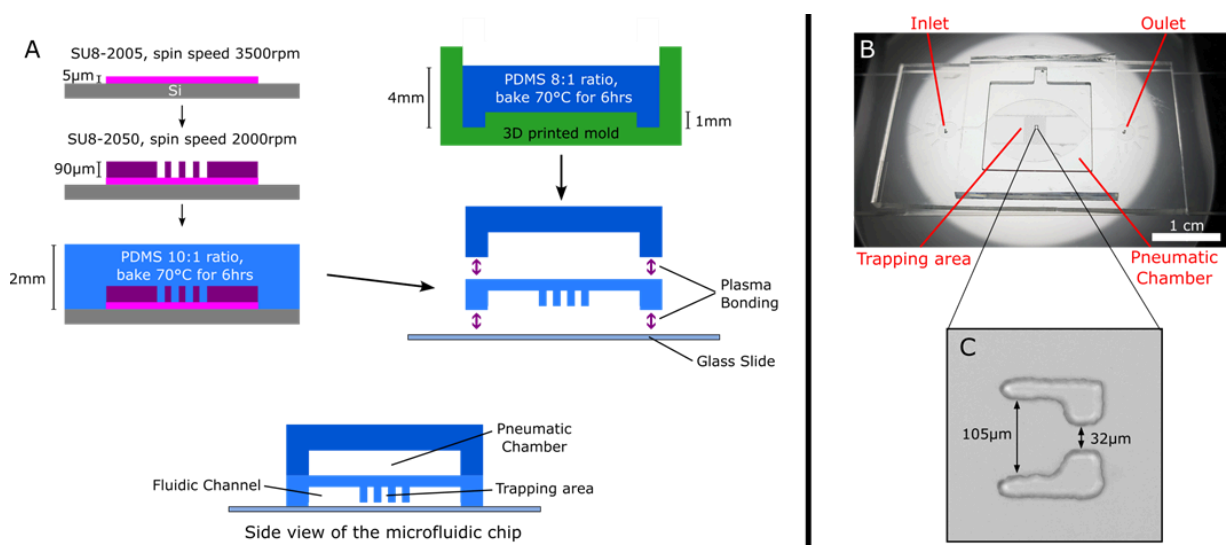

*Figure 2 – Fabrication and Structure of the Microfluidic Chip: (A) Sequential fabrication steps using photolithography. Two photolithography layers define a channel with a raised trap array relative to the glass substrate onto which the PDMS component is bonded. A second PDMS layer, molded from a 3D-printed structure, forms the pneumatic chamber. Together, these components create a chip with both a pneumatic control layer and a particle-trapping microchannel. (B) Photograph of the assembled microfluidic chip. (C) Close-up of an individual trap within the array, designed to capture particles ranging from 50 to 90 µm.*

The microfluidic channel and trap array are fabricated using two layers of negative photoresist (SU-8 2005 and SU-8 2075, MicroChem Corp.), spin-coated at 4000 rpm and 2000 rpm for 30s, yielding thicknesses of 5 µm and 90 µm, respectively. The first layer is set for the elevation of the second layer : the trapping array will be elevated and not be bonded to the glass surface. The cavity geometry is optimized to ensure reliable lifting of the trap array under negative pressure (See Section 3.1).

PDMS (RTV 615, GE Silicones) is mixed at a 10:1 base-to-curing-agent ratio, poured onto the master mold, degassed under vacuum for 30min, and cured at 70 °C for 12 hours. In parallel, the pneumatic chamber mold is produced via 3D printing (Mars 2 Pro, Elegoo) using standard photopolymer resin (Elegoo). The printed mold is post-cured in an oven at 70 °C for 3 days to ensure complete cross-linking. This mold is used to cast the second PDMS structure, using an 5:1 ratio to increase rigidity. The PDMS is again degassed and cured under identical conditions.

After curing, both PDMS components are peeled off and trimmed. They are bonded together via oxygen plasma treatment (CUTE Vacuum Plasma System, Femto Science). The entire structure is then plasma-treated and carefully bonded to a glass slide (VWR, 26 × 76 mm) to prevent the trapping array to touch the glass. To further reinforce the pneumatic chamber, an additional glass coverslip can be bonded to the top surface.

### 2.3. Experimental Setup and Fluid Handling

After fabrication, the microfluidic chip is placed at the center of a custom setup enabling both fluidic actuation and real-time observation. The pneumatic chamber is connected to a syringe pump (AL-4000, World Precision Instruments), allowing the application of positive or negative pressure within the pneumatic cavity. The fluidic inlet of the chip is connected to a pressure controller (MFCS™ series, Fluigent).

One unique solution was used to demonstrate proof-of-concept functionality and potential applications of the system. Indeed, an aqueous solution (DI water, Millipore, 18.2 MΩ.cm) with 10% w/w Poloxamer 188 (CAS no. 9003-11-6, Sigma-Aldrich) was prepared. To create an isodense medium for the particles, 7.5% w/w sodium chloride (NaCl, CAS no. 7647-14-5, Sigma-Aldrich) was added.

Polystyrene microspheres (80 µm diameter, Thermo Scientific™) were fluorescently labeled using Nile Red (CAS no. 7385-67-3, Sigma-Aldrich) . A stock solution of Nile Red was prepared by dissolving 1 mg of Nile Red powder in 1 mL of absolute acetone (CAS no. 67-64-1, VWR) and vortexing gently until fully dissolved. A 1 mL aliquot of the microsphere suspension was transferred into a centrifuge tube and centrifuged at 5000 g for 2 minutes. The supernatant was discarded, and the pellet was resuspended in 1 mL of DI water. For fluorescent labeling, 20 µL of the Nile Red stock solution was added to the microsphere suspension. The mixture was vortexed and incubated for 30 minutes at room temperature in the dark to prevent photobleaching. After incubation, the microspheres were centrifuged again to remove excess dye, and the supernatant was discarded. The pellet was washed three times with ultrapure water and finally resuspended in the previously prepared NaCl solution. The labeled microspheres were stored at 4 °C in the dark until further use.

### 2.4. Measurement of Trapping Membrane Deformations by Confocal Microscopy

To quantify the deformation of the trap array under applied pressure, the microfluidic chip is filled with ethanol (CAS no. 64-17-5, VWR). A Leica SP8 confocal microscope equipped with a 10× dry objective is used in bright-field mode to focus on specific reference features within the array. Positional differences between these features are recorded to assess the vertical deformation of the trap array.

Pressure is precisely controlled using the LineUp™ Push-Pull system (Fluigent), with values set at [0, −20, −50, −80, −100, −120] mbar. At each pressure level, the focus is adjusted to the reference features. Error bars correspond to the focal range within which the precise focus becomes ambiguous. As the applied pressure increases, the deformation causes the trap array to deviate from a single focal plane, resulting in broader error margins. The measured uncertainties are [2, 4, 6, 8, 10, 12] µm for the respective pressure levels.

Due to the use of a non-immersed objective, deformation measurements along the z-axis must be corrected for optical path distortion. Light travels through a sequence of media—air, glass, and ethanol—each with a different refractive index, causing apparent z-position errors. To compensate, the measured displacement values are corrected using:

$$\Delta Z_{real} = \frac{\Delta Z_{measured}}{n_{effective}}$$

where $n_{effective}$ is the effective refractive index for the full optical path, computed as:

$$n_{effective} = \frac{\sum_i n_i t_i}{\sum_i t_i}$$

with $n_i$ the refractive index and $t_i$ the thickness of each medium *i*. Based on the known indices and thicknesses of air, glass, and ethanol in the optical path, we applied a correction factor of $n_{effective}$ = *1.05*, leading to:

$$\Delta Z_{real} = \frac{\Delta Z_{measured}}{1.05}$$

### 2.5. Microscope and fluidic control

The particle trapping and releasing is controlled using a Micro-Manager plugin we developed [18]. Micro-Manager is a free, open-source platform to control microscopes, which recently got extended to also control fluidic components (like syringe pumps and pressure controllers). We can, therefore, control the entire experiment from a single application.

The experiment consists of three phases: Trapping, Imaging, and Releasing. The plugin loads a predefined "Preset" at the start of the Trapping and Releasing phase. A "Preset" is Micro-Manager's way to represent a grouped collection of settings, like active objective, active filter-cube, and illumination strength, that can be changed collectively. This enables us to activate/deactivate the traps, change the illumination settings, and move to a low magnification objective to observe as many traps as possible during the Trapping and Releasing phase.

The Imaging phase works slightly differently. We take full advantage of Micro-Manager's powerful Multi-Dimensional Acquisition (MDA), which allows us to predefine a list of positions and a list of presets, after which Micro-Manager explores the entire parameter-space, taking a picture at each position for each preset. It is, therefore, possible to take multiple images using multiple filter sets and illumination sources to perform, for instance, correlations between fluorophores for each of the trapped objects.

## 2.6. Image analysis

We've developed an image analysis pipeline to count the number of filled traps, to trigger the MDA. The image analysis is a four-step process. First, the image is inverted, such that the dark edge of the particles become white, against a black background. Next, a mean blur is applied to remove some of the noise generated by the camera. Next, the image is thresholded to remove the background. Lastly, the holes in the particles are filled, to yield solid particles. The particles are then counted using a connected component algorithm.

## 3. Results and Discussion

### 3.1. Design of Microfluidic Trapping Chambers

The microfluidic design was inspired by the work of O. Mesdjian et al. (2019) [8], focusing on the development of a Pachinko-type trap array. This design features a sequence of 15 columns, each containing 14 traps, resulting in a array of 210 traps. As illustrated in Figure 2C, each trap consists of a cavity (105 × 105 µm²) to house particles and an orifice (32 µm) allowing fluid flow. Particles are fluidically directed into the traps but remain confined within the microstructures, defining the core trapping mechanism of the Pachinko system.

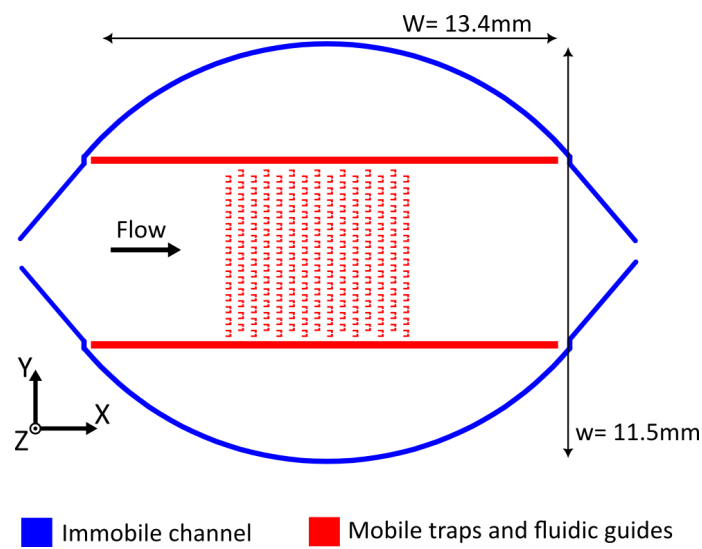

*Figure 3. Schematic of the cavity containing the trap array. The figure highlights the regions that remain fixed and those that are mobile along the z-axis. The overall structure exhibits a cylindrical shape, with two guiding features designed to direct the flow and particles toward the traps.*

However, in conventional Pachinko-like arrays, the channel walls guiding particles toward the trapping zone are rigid [8], with no pneumatic cavity. Consequently, when deformation is applied along the z-axis, minimal displacement occurs near the traps adjacent to the channel edges, which prevents particle release. To overcome this limitation, we designed a wider, elliptical cavity to maximize the distance of the trapping zone from the cavity edges. Additionally, two rectilinear guiding structures were integrated to mimic the role of rigid boundaries in the classical Pachinko design (Fig. 3). Using our custom fabrication protocol (see Section 2.2), these guides—as well as the traps themselves—were elevated above the glass surface, making them vertically mobile. This configuration ensures that particles are directed toward the trap array and not into the two elliptical lobes, where the flow remains low due to narrow entrances.

### 3.2. Theoretical Basis for Membrane Deformation

To understand how membrane deformation enables particle release, theoretical considerations are required. Gervais et al. (2006) [19] proposed a model for PDMS deformation under fluidic pressure in microfluidic channels. For a channel with width *W* and height *h*, a scaling approximation based on Hooke's law is used (Eq.1), where *E* is the Young's modulus of PDMS and *σ* is the applied stress.

$$\varepsilon = \frac{\sigma}{E} \quad (1)$$

Two displacement components are considered: vertical and horizontal (Eq.2 and Eq.3).

$$\varepsilon_{vertical} \sim \frac{\Delta h}{W} \quad (2)$$

$$\varepsilon_{lateral} \sim \frac{\Delta W}{h} \quad (3)$$

Given that W>>h in our setup, only vertical displacement is significant, leading to:

$$\Delta h_{max} = c_1 \frac{W}{E} P \quad (4)$$

Where $c_1$ is a proportionality constant which depends on the channel geometry and mechanical properties. *P* is the pressure applied to the surface and causing the deformation.

Unlike Gervais' article [19], which relates pressure to internal fluid flow within the channel, we directly impose pressure using a pneumatic cavity positioned above the channel, and we measure it independently. Since the model from Gervais only assumes a semi-infinite PDMS layer (>2 mm thick), we refined it for finite PDMS thickness as described by Raj et al. (2017) [20].

For small deformations, we adopt the parabolic profile for the membrane as described by Schomburg et al. (2015) [21], leading to the "thick membrane" approximation (Eq.5), where $t$ is the membrane thickness (100 µm < $t$ < 2 mm), and $\nu$ is the Poisson's ratio of PDMS.

$$\Delta h_{max} = \frac{W^4(1-\nu^2)}{66\, t^3 E} P \quad (5)$$

For thinner membranes ($t$ <100 µm), a different scaling behavior is observed [22], and a power law governs the relationship between pressure and deformation, given by (Eq.6).

$$\Delta h_{max} = \left(\frac{3W^4(1-\nu)}{64\, t\, E} P\right)^{1/3} \quad (6)$$

These models provide the theoretical framework to understand the deformation behavior of the trap array. In our study, we conducted deformation measurements to determine which model best fits our microfluidic system.

### 3.3. Experimental results for the deformation

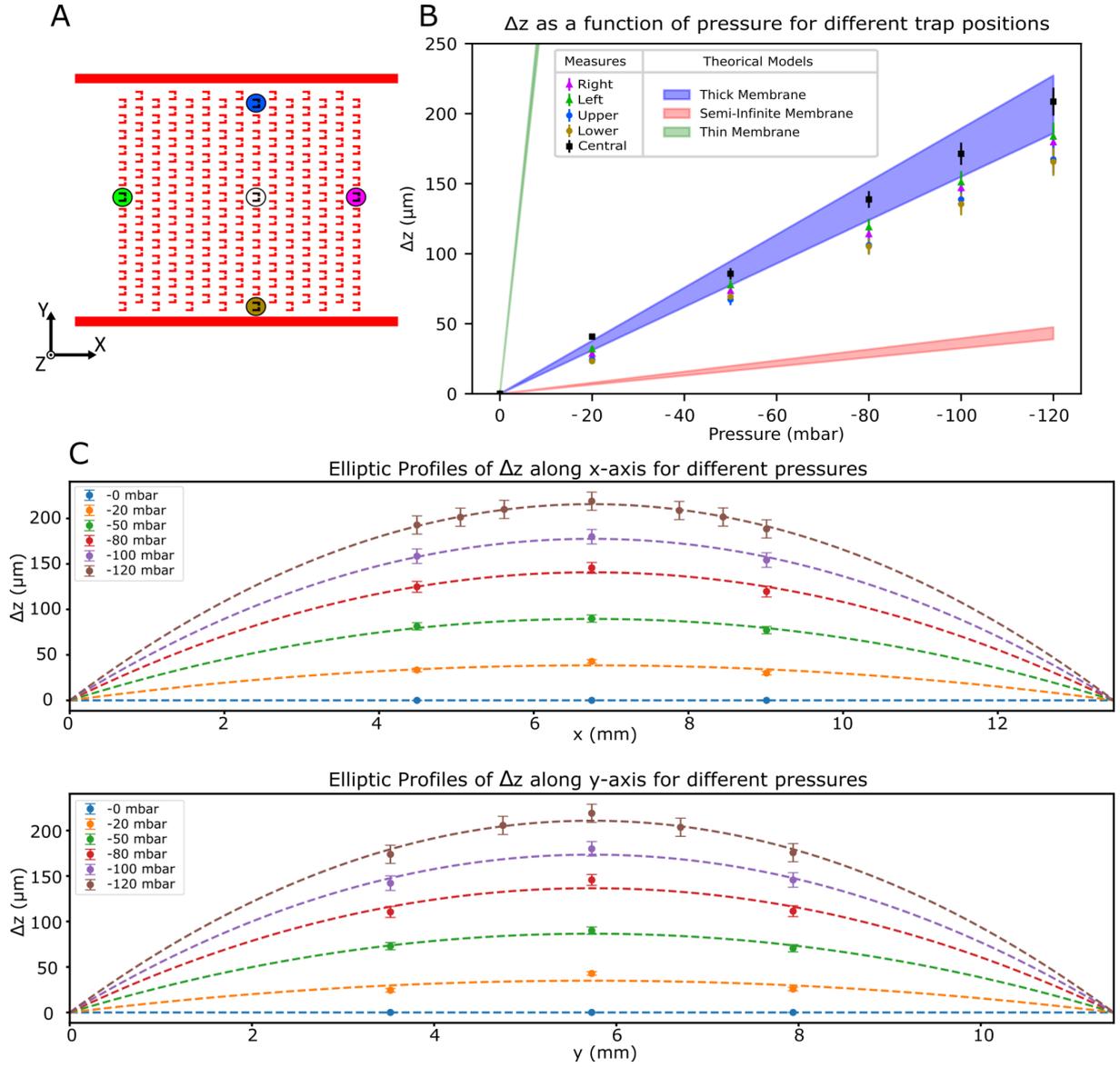

*Figure 4. Study of the deformations of the trapping array (A) Schematic of the trap array indicating the four observation points selected for the deformation study. (B) Graph showing experimental deformation as a function of applied negative pressure, overlaid with theoretical predictions from Eq. 4, Eq. 5, and Eq. 6. Model parameters: membrane thickness t =1.8mm; membrane width w =11.5 mm; empirical constant $c_1$ = 0.62 [23]; Poisson's ratio $\nu$ =0.5 [24]; and PDMS Young's modulus E=2±0.2 MPa [25]. (C) Deformation profiles along the x- and y-axes under increasing negative pressure in the pneumatic cavity. The experimental data are fitted with parabolic fits with limit conditions at the edges of the cavity ($\Delta z(x=0) = \Delta z(xmax) = 0$ and $\Delta z(y=0) = \Delta z(ymax) = 0$).*

To assess the release capability of trapped particles, we performed deformation measurements of the membrane (See Section 2.4). The results are presented in Figure 4.

By applying negative pressure through the pneumatic cavity, we induced controlled deformation of the membrane. Z-position measurements were collected at multiple trap locations using confocal microscopy (**Fig. 4A).** As shown in **Fig. 4B**, deformation increased proportionally with the magnitude of the applied negative pressure. Notably, the central traps experienced greater displacement compared to those located at the periphery of the array.

To further interpret the results, we compared the experimental data with theoretical deformation models (see Section 3.2). For these models, we measured the thickness of the membrane containing the traps ($t$ = 1.8 mm) and the width of the deformed region ($w$ = 11.5 mm). Additionally, literature values were used for parameters that could not be directly measured: the coefficient $c_1$ = 0.62 [23] and the Poisson's ratio $v$ = 0.5 [24] and the Young's modulus of PDMS was taken as $E$ = 2 ± 0.2 MPa [25].

Among the models considered, the thick membrane model (Eq.5) from Raj et al. [20] aligned most closely with the experimental data, confirming its relevance for our configuration.

**Fig.4C** presents deformation profiles along the x- and y-axes of the trap array. These profiles help define the critical pressure required to release trapped objects. In our setup, negative pressures exceeding –100 mbar resulted in a membrane elevation of at least 120 µm, which is sufficient to release particles with a diameter of 80 µm.

These findings confirm that the fabrication and actuation protocol reliably enables pressure-induced particle release from the trap array.

### 3.4. Trapping Process

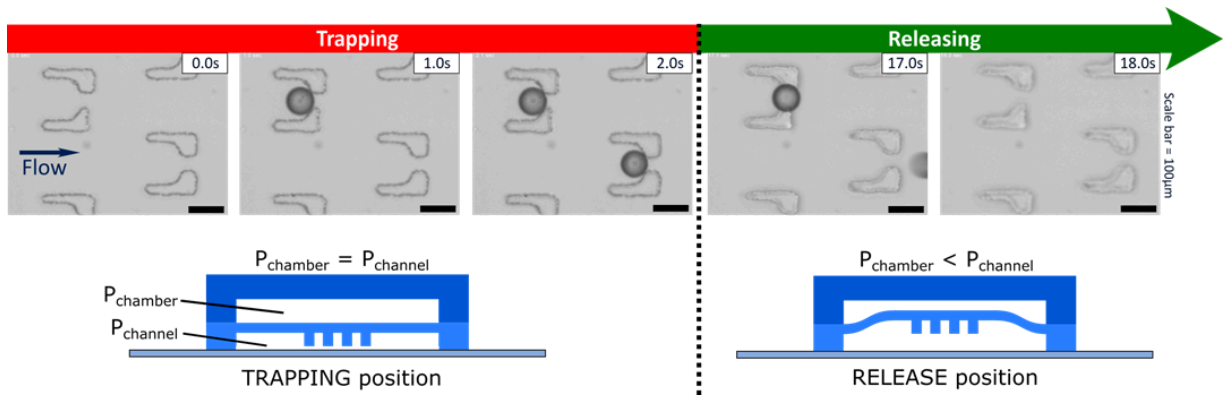

*Figure 5. Trapping and subsequent release process of 80 µm diameter polystyrene particles in a solution of water with 10% Poloxamer 188. 0.0sec: The traps are initially empty and pressed against the surface of the glass chip ($P_{chamber} \geq P_{channel}$). 1.0-2.0sec: First one particle, then a second, are trapped using flow control and the traps. 17.0sec: The pressure in the chamber is gradually reduced, and the beads begin to escape ($P_{chamber} < P_{channel}$). 18.0sec: Finally, the beads are completely released as the traps are now out of focus, pulled above the surface by the pneumatic chamber.*

Once the setup is positioned under the microscope and connected to the fluidic and pneumatic actuation systems, particle trapping and release can start. For demonstration purposes, we focused on two central traps within the array (**Fig. 5**). A pressure of 100 mbar was maintained in the solution reservoir, and the syringe pump applied an overpressure to the pneumatic chamber to ensure traps were in the capture position ($P_{chamber} \geq P_{channel}$). Within seconds, the first and then the second trap filled with an 80-µm-diameter particle, with the entire trapping process completed in a matter of seconds.

The release phase involved increasing the flow rate in the fluidic channel by applying a pressure of 150 mbar. Subsequently, a vacuum was induced in the pneumatic cavity via the syringe pump ($P_{chamber} < P_{channel}$), generally ensuring the traps moved out of the microscope's focal plane. Gradually, the particles followed the fluid flow and exited the traps toward the outlet.

## 3.5. Application of the fully-automated technology

To illustrate the applicability of the developed technology, we conducted an experiment using the particle solution previously prepared (see Section 3.2). This case study bridges the physical principles validated in the proof-of-concept with the automation features introduced in this work. The user interface—detailed in the Methods section—plays a central role in coordinating the various components required for full automation of the system.

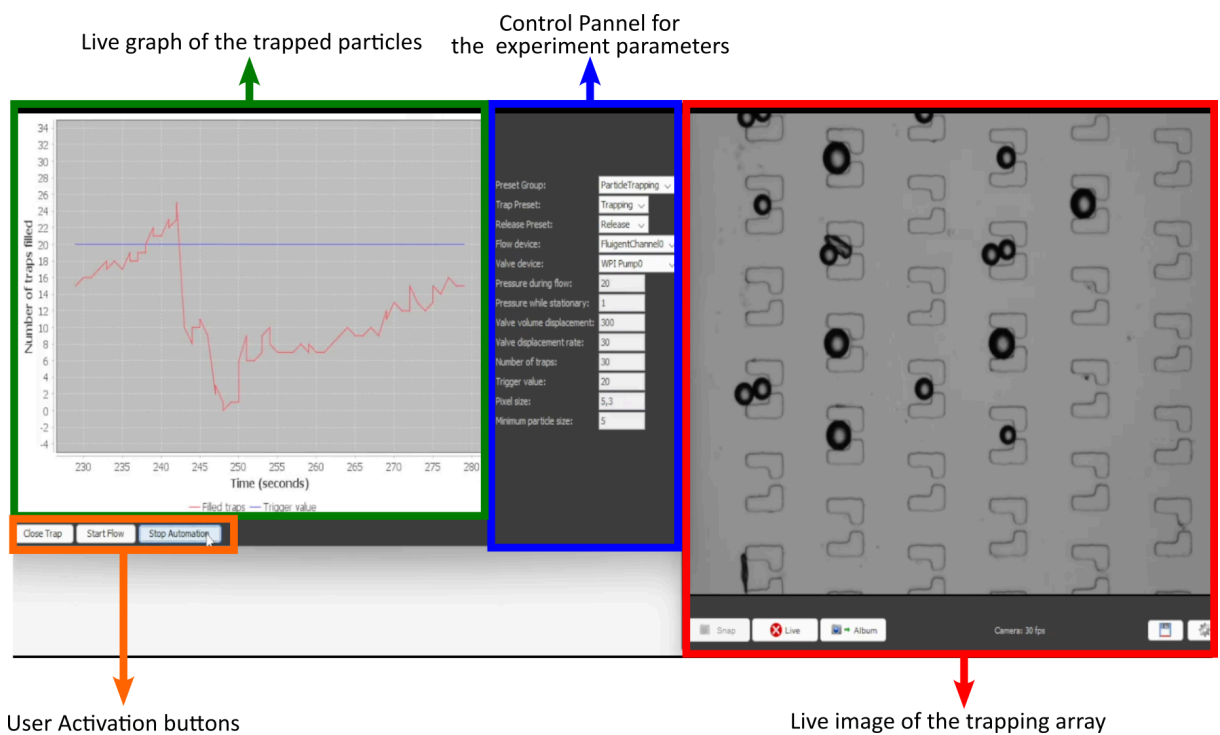

*Figure 6. Screenshot of the interface designed for the trapping/releasing experiments.*

This interface, combined with the µManager libraries, was specifically designed to control the XY motorized stage, microscope objectives, optical filters, as well as the fluidic and pneumatic systems. Several key functions were optimized to streamline tasks such as particle trapping, imaging, and release. A dedicated preprocess is set to detect the positions of the traps (**Fig.7A**). The user simply inputs three key points on the chip, enabling the interface to automatically determine the coordinates of the trap array for subsequent operations. Following this brief calibration step, the entire experimental cycle—comprising trapping, image acquisition, and particle release—can be executed autonomously.

In our case study, we successfully performed complete cycles of trapping and releasing under bright field observation. The graph (**Fig.7B**) highlights the trapping and the liberation of the particles during an experiment. Detection of particles in the traps has to be improved to avoid the noise that appear after the liberation of the particles. Indeed, the ROIs are influenced by the particles leaving the trapping array. The release of particles is there manually induced, but the interface can allow the user to set a trigger value for the liberation of particles. During each sequence, the system is built to acquired high-resolution single-cell images of individual traps (**Fig.7C**).

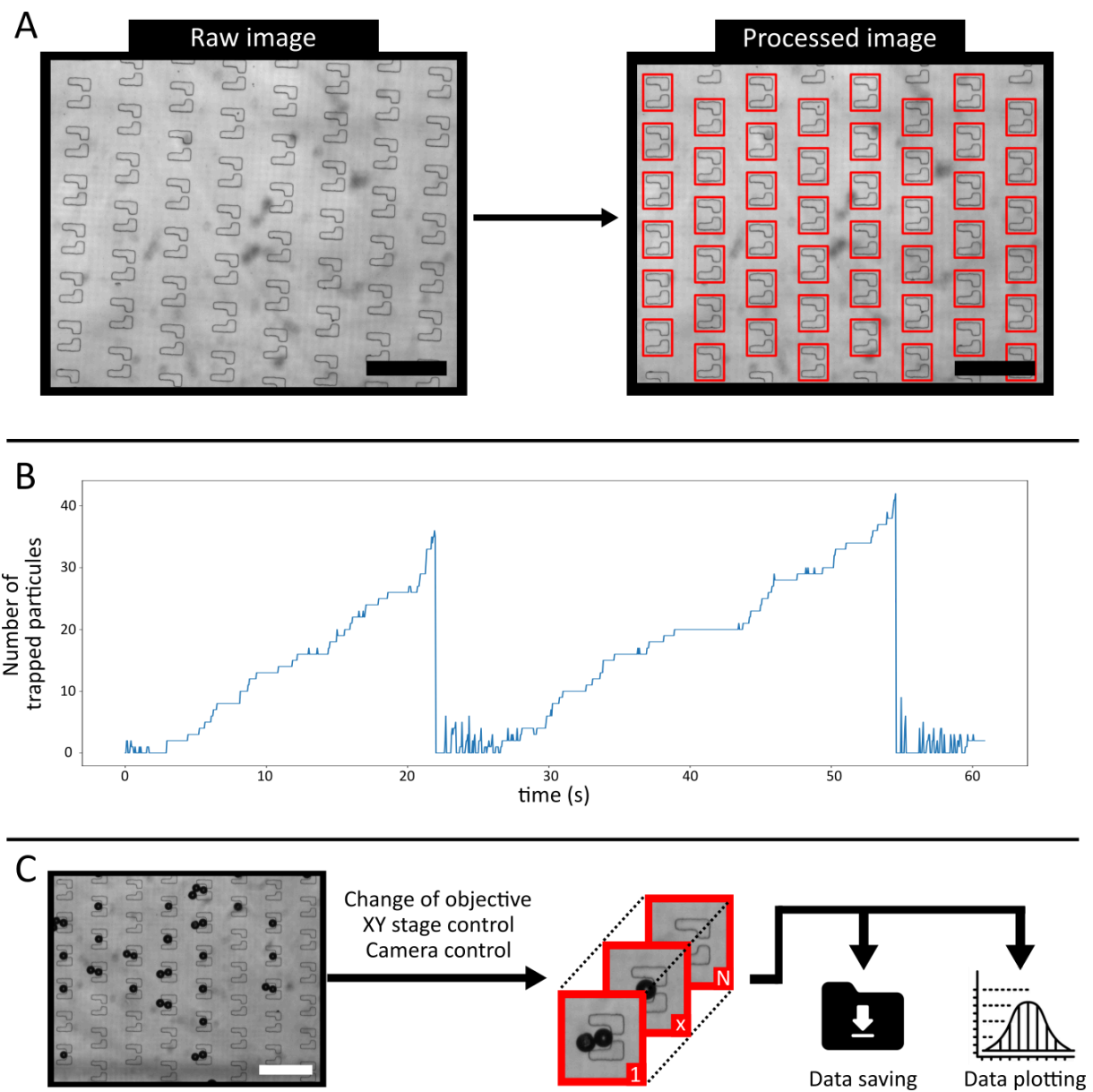

*Figure 7. Fully-automated process using the 80 µm-wide particles solution. (A) Preprocess for live detection of the traps and rotation of the image. Scale bar : 500 µm. (B) Graph showing the number of trapped particles over time, highlighting the trapping and release phases across the full array. (C) Analysis Process of the trapping array. The interface can allow for single-trap analysis of particles. Scale bar : 500 µm.*

**Conclusion and Perspectives**

The developed device supports continuous trapping and release experiments by integrating a Pachinko-style trap architecture with PDMS membrane deformation and pneumatic valve principles. This combination enables successive and reversible particle immobilization within a single platform. The system's functionality was validated using traps and microparticles approximately 80 µm in diameter. While effective, this size range is relatively large for typical microfluidic applications. A key direction for improvement is the miniaturization of trap dimensions to accommodate smaller biological elements, such as single cells, which are typically around 10 µm in size. Additionally, optimizing the spatial arrangement of the traps [8] could further enhance trapping efficiency and throughput.

**Acknowledgments**

This work was supported by the Agence Nationale de la Recherche (ANR) under grant number ANR-21-CE43-0021. This work benefited from the technical contribution of the joint service unit CNRS UAR 3750. The authors would like to thank the engineers of this unit (and in particular Taha Messelmani, Izadora Fujinami Tanimoto and Elian Martin) for their advice during the development of the experiments.

**Author contribution**

AC: Investigation, Methodology, Validation, Visualization, Writing - original draft, Writing - reviewing and editing.
LK: Resources, Writing - original draft, Formal Analysis, Methodology.
JF: Conceptualization, Methodology, Funding acquisition, Supervision, Visualization, Writing - original draft, Writing - reviewing and editing.

**Competing interests**

The authors declare that they have no competing interests.

**Data and materials availability**

Data and materials will be made available on request.